\documentclass[aps,prb,twocolumn,superscriptaddress,showpacs,amsmath,amssymb,longbibliography]{revtex4-2}
\usepackage[english]{babel}
\usepackage{amsmath}
\usepackage{bm}
\usepackage{graphicx,bbm} 
\usepackage{times}
\usepackage{epsfig} 
\usepackage[utf8]{inputenc}
\usepackage[colorlinks,linkcolor=blue,citecolor=blue,urlcolor=blue]{hyperref}
\usepackage{color}
\usepackage{latexsym}
\usepackage{ulem}
\definecolor{nred} {RGB}{224,0,0}
\definecolor{nblue} {RGB}{28,130,185}
\definecolor{dgreen} {RGB}{78,138,21}
\definecolor{norange}{RGB}{230,120,20}

\begin{document} 
\title{Many-body localization as a percolation phenomenon}
\author{P. Prelov\v{s}ek}
\affiliation{Jo\v zef Stefan Institute, SI-1000 Ljubljana, Slovenia}
\affiliation{Faculty of Mathematics and Physics, University of Ljubljana, SI-1000 
Ljubljana, Slovenia}
\author{M. Mierzejewski}
\affiliation{Department of Theoretical Physics, Faculty of Fundamental Problems of Technology, 
Wroc\l aw University of Science and Technology, 50-370 Wroc\l aw, Poland}
\author{J. Krsnik}
\affiliation{Institute of Physics, HR-10000 Zagreb, Croatia}
\author{O. S. Bari\v si\'c}
\affiliation{Institute of Physics, HR-10000 Zagreb, Croatia}

\date{\today}
\begin{abstract}
We examine the standard model of many-body localization (MBL), i.e., the disordered chain of interacting spinless fermions, by representing it as the network in the many-body (MB) basis of noninteracting localized Anderson states. By studying eigenstates of the full Hamiltonian, for strong disorders we find that the dynamics is confined up to very long times to disconnected MB clusters in the Fock space. By keeping only resonant contributions and simplifying the quantum problem to rate equations (REs) for MB states, in analogy with percolation problems, the MBL transition is located via the universal cluster distribution and the emergence of the macroscopic cluster. On the ergodic side, our approximate RE approach to the relaxation processes captures well the diffusion transport, as found for the full quantum model. In a broad transient regime, we find an anomalous, i.e., subdiffusivelike, transport, emerging from weak links between MB states.  

\end{abstract}

\maketitle


\section {Introduction} 

The interplay of disorder and interaction in dense fermionic systems remains a challenging theoretical problem, offering a novel scenario for a possible metal-insulator transition at elevated temperatures $T > 0$. While the (Anderson) localization of noninteracting fermions \cite{anderson58} is a well-established phenomenon in disordered systems, the role of interaction has been seriously theoretically addressed only within the last 15 years \cite{basko06,oganesyan07}. Based mostly on numerical studies, the transition/crossover towards the many-body localization (MBL) has been revealed by several criteria: change in level statistics \cite{oganesyan07,torres15,luitz15,serbyn16}, slow growth of entanglement entropy\cite{znidaric08,bardarson12,serbyn15}, vanishing d.c. transport \cite{barisic10,berkelbach10,agarwal15,gopal15,barlev15,steinigeweg16,prelovsek17}, and the absence of thermalization \cite{pal10,luitz16,mierzejewski16}, where all these criteria may be related to the existence of a macroscopic number of integrals of motion in the MBL phase \cite{serbyn13,huse14,ros15,rademaker16,mierzejewski18,Roeck2017}. Importantly, qualitative features of the MBL have been found in several cold-atom experiments \cite{schreiber15,kondov15,bordia16,luschen17}.

While model investigations of strongly disordered systems confirm the existence of the MBL, the phenomenology of the transient regime towards the normal diffusion is much less clear, making the physical interpretation of the interplay between interaction and strong disorder difficult. For the transient regime, there is abundant evidence of a subdiffusive nature of transport \cite{agarwal15,gopal15,gopal16,znidaric16,schulz20}, accompanied by large and anomalous fluctuations of relevant quantum observables near the MBL transition \cite{barisic16,mierzejewski20}. However, the existence of this transient regime and its origin is still under dispute \cite{bera2017,suntajs19,Bera2019,Sierant2020}. As a plausible explanation, Griffith effects due to weak-link structures in real space have been proposed \cite{agarwal15,gopal15,gopal16,luitz16}. In recent works, there is growing numerical evidence emphasizing the percolation aspects of the MBL transition, whereby the multifractal clustering appears in the many-body (MB) Fock space \cite{serbyn15,serbyn17,roy191,roy192,mace19}, with implications for both sides of the MBL transition.

We present numerical results for the prototype quantum MBL model that clearly reveals disorder-induced clustering effects. More precisely, we show that the basis consisting of localized MB states tends to split into disconnected clusters such that the dynamics of the system is confined over very long times to MB clusters. These effects become even more evident in the formulation of the problem when quantum transitions are replaced by rate equations (REs). Such a RE approach was recently applied by the present authors to several quantum models of the propagation of (single) particles coupled to bosons \cite{mierzejewski19,mierzejewski20}, revealing normal as well as subdiffusive transport behaviors. In the present case, it allows for a direct study of thermalization of MB charge density. RE results are consistent with full quantum calculations, confirming nearly exponential decrease of diffusion with disorder strength. In the transient regime that spans a broad range of parameters below the MBL transition, we find anomalous/subdiffusive relaxations towards equilibrium, which can be attributed to weak links between MB states, with no apparent relation to the real-space weak links.

\section {Model, resonance condition, and rate equations} 

We consider the prototype model of MBL, with interacting spinless fermions in the 
disordered one-dimensional system, equivalent to the XXZ spin-$1/2$ chain with random fields
\begin{equation}
H = - t_h \sum_{i} \left( c^\dagger_{i+1} c_i + \mathrm{H.c.}\right)
+ \sum_i \epsilon_i n_i\ + V \sum_i n_{i+1} n_{i} , \label{tvh}
\end{equation}
\noindent where the quenched local disorder is characterized by a uniform distribution $-W < \epsilon_i <W$. By setting $V=2t_h$, we focus on the most frequently studied case of the isotropic Heisenberg chain, for which the MBL transition/crossover is expected at $W_c \sim 6 -8$ \cite{luitz15,barlev15}. However, it should be noted that significantly larger values of $W_c$ were reported recently as well \cite{chanda2020,Devakul2015,Doggen2018,
Bera2019,Panda2020}. We use $t_h=1$ as the unit of energy and focus on the half-filled case with $N=L/2$ fermions, where $L$ is the number of lattice sites. 

It is convenient to represent the model, Eq.~(\ref{tvh}), in terms of Anderson single-particle states $\phi_{li}$, in order to obtain the noninteracting part of the Hamiltonian in the diagonal form. The latter is given by the single-particle occupation numbers, $\tilde H_0 = \sum_l \epsilon_l  \varphi^\dagger_l \varphi_l$, with $\varphi^\dagger_l   = \sum_i \phi_{li} c_i^\dagger $.  In particular, within the (localized) MB basis, $|\underline{n} \rangle = \prod_l (\varphi^\dagger_l)^{n_l} | 0 \rangle $, the interaction $V$ term may be written \cite{prelovsek18,laflorencie20} in terms involving two, three, and four single-particle operators, respectively,

\begin{eqnarray}
H^\prime_2 &=& \sum_{k>l} h^2_{lk} n_l n_k, \qquad H^\prime_3 =  \sum_{j \neq k \neq m } 
h^3_{jkm} n_j \varphi_m^\dagger \varphi_k,\nonumber \\
H^\prime_4 &=&  \sum_{(j > k) \ne (m > l) } h^4_{jklm}  \varphi_l^\dagger \varphi_m^\dagger \varphi_k \varphi_j,
\end{eqnarray}

\noindent where  $h^3_{jkm} = h^4_{jkjm} $, $h^2_{jk} = h^4_{jkjk} $, and 

\begin{eqnarray}
h^4_{jklm} &=& V( \chi_{jk}^{lm} + \chi^{jk}_{lm} - \chi_{kj}^{lm} -\chi_{jk}^{ml} ) \nonumber \\
\chi_{jk}^{lm} &=& \sum_i  \phi^*_{li} \phi^*_{m,i+1} \phi_{k,i+1} \phi_{j,i}\;.\label{chi}
\end{eqnarray} 

\noindent The Hartree-Fock term $H'_2$ is diagonal and may be included in the unperturbed Hamiltonian $H_0=\tilde{H}_0
+ H_2'$. This diagonal part of the Hamiltonian defines the energies of the MB basis states, $E^0_{\underline n} = \sum_l n_l \epsilon_l + \sum_{k>l} h^2_{lk} n_l n_k$, while the remaining non-diagonal part, given by $H^\prime = H'_3 + H'_4$, mixes different states $| \underline n \rangle$, allowing for charge fluctuations and the charge delocalization.

When discussing the dynamics and real transitions between MB states inside or in the vicinity of MBL regime (as opposed to virtual ones, representing perturbative corrections), one should primarily consider those matrix elements between the MB states that satisfy the resonance condition (RC),
\begin{equation}
|H'_{\underline n \underline n'}| > R ~ (E^0_{\underline n}-E^0_{\underline n'}), \qquad 
H'_{\underline n \underline n'} = \langle \underline n| H' | \underline n' \rangle\;, \label{rc}
\end{equation}
with the RC parameter $R \lesssim 0.5$. The same reasoning is used for the reduced-basis approach \cite{prelovsek18,laflorencie20}.

Taking a step further by neglecting the quantum coherence of hopping between MB states, one can regard the relaxation from some initial MB occupation profile towards the equilibrium as a cascade of (irreversible) transitions, described by the RE for the MB occupations/probabilities, $p_{\underline n}$, 
\begin{equation}
\frac{d}{dt} p_{\underline n} = \sum_{\underline n' \ne \underline n} 
\Gamma_{\underline n \underline n'} (p_{\underline n'} - p_{\underline n}) = \sum_{\underline n'} 
\tilde \Gamma_{\underline n \underline n'} p_{\underline n'}, \label{re}
\end{equation}
satisfying $\sum_{\underline n} p_{\underline n} = 1$. Transition rates $\Gamma_{\underline n \underline n'} > 0$ take a Fermi's golden rule form, $\Gamma_{\underline n \underline n'} = \zeta |H'_{\underline n \underline n'} |^2 {\cal N}_{\underline n \underline n'}$, with the MB density of states that is relevant to the transition,
\begin{equation}
{\cal N}_{\underline n \underline n'} \sim \frac{R}{2 |H'_{\underline n \underline n'}| }
\Theta( \frac{ |H'_{\underline n \underline n'}|}{R}
- |\Delta E^0_{\underline n \underline n'} |)\;. \label{nnn}
\end{equation}
Although $\zeta$ does not influence any qualitative conclusions, to set the time scale in units corresponding to oscillations of two-level systems, we choose $\zeta=1/(\pi R)$. On the other hand, the choice of $R$ directly influences the results, in particular the MBL transition value $W_c$. While the conservative RC would be $R=0.5$, we use a bit softer cutoff, $R =0.3$.

\section{Many-body clustering and the percolation threshold} 

\begin{figure}[tb]
\includegraphics[width=0.8 \columnwidth]{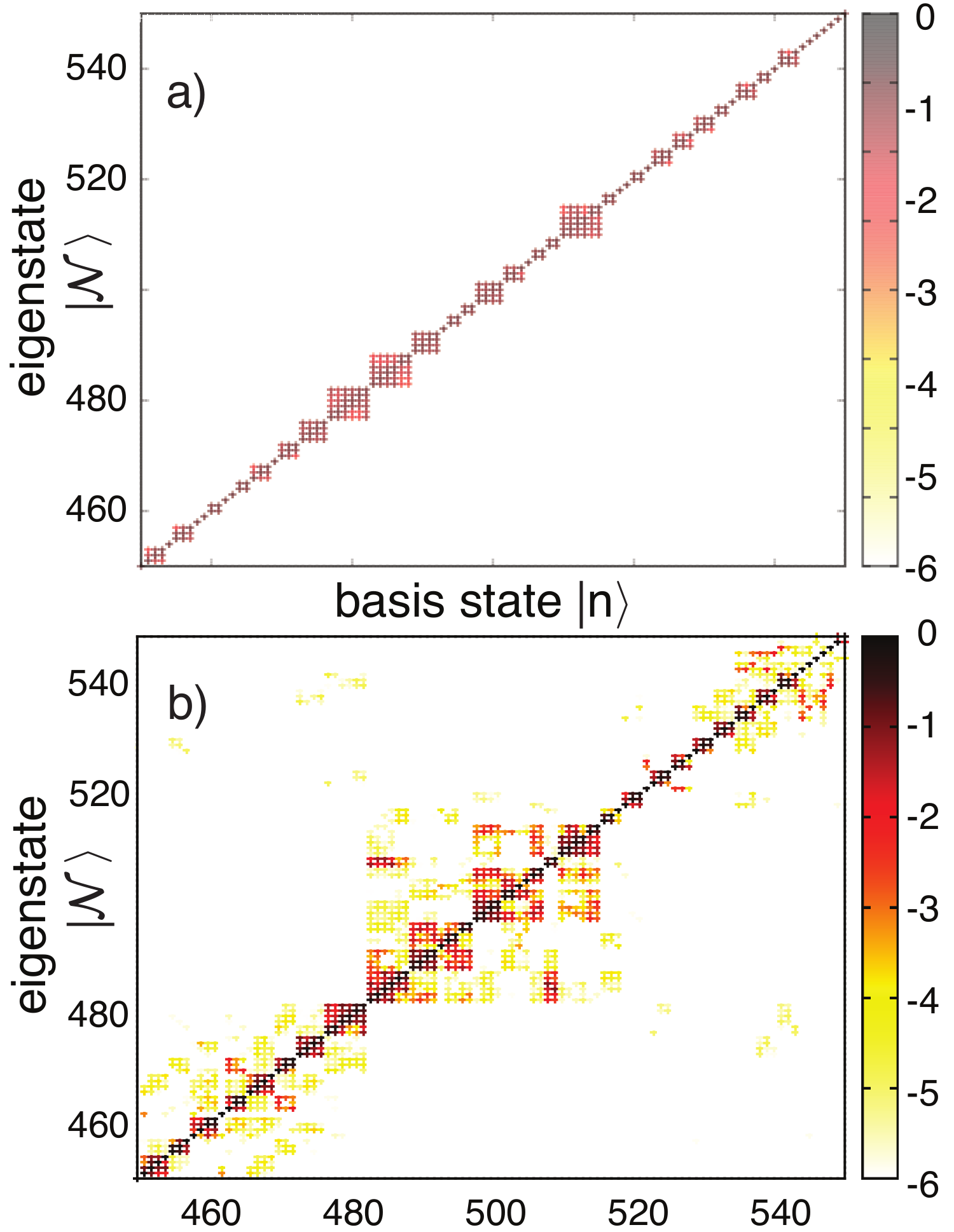}
\caption{Clustering of eigenstates for a) the truncated and b) the full Hamiltonians for disorder $W=10$ on the $L=12$ system. 
Colors are assigned to the numerical values of $\log_{10}|M_{\underline n,{ \cal N}}|^2 $. }
\label{fig1}
\end{figure}

Perfect clustering means that  the Anderson MB basis, $| \underline n \rangle $,  may be split into disconnected sets (clusters) such that each MB eigenstate $| {\cal N} \rangle$ has a projection on only a single cluster. 
Such clustering  can be univocally identified when  $M_{\underline n,{ \cal N}}=\langle \underline n | {\cal N} \rangle$  is a block-diagonal matrix, whereby each block corresponds to  a separate cluster.
However, the eigenstates are typically sorted according to the energies, whereas the numbering of the MB basis is arbitrary. 
Therefore, in order to reveal such a block-diagonal structure, the rows and columns of   $M_{\underline n,{ \cal N}}$ should be permuted accordingly.
 To this end, we apply a simple sorting algorithm, based on an observation that for the eigenstates $| {\cal N} \rangle$ and $| {\cal N}' \rangle$, which belong to different clusters,
\begin{equation}
O({\cal N},{\cal N}' )=\sum_{\underline n} |\langle \underline n | {\cal N}\rangle| |\langle \underline n | {\cal N}' \rangle|\;, 
\end{equation} 
vanishes. Starting from an arbitrary $| {\cal N} =1 \rangle$, we find ${\rm max}_{{\cal N'}>{\cal N}} O({\cal N},{\cal N}' ) $, and the resulting 
$| {\cal N'} \rangle$ is then considered a consecutive  eigenstate $| {\cal N} =2 \rangle$. The same procedure is repeated for each newly added state $| {\cal N} \rangle$, grouping together, in this way, the eigenstates belonging to the same cluster. In the second stage, we sort the basis states $ | \underline n \rangle $ according to their projections on already sorted eigenstates. That is, for each $| {\cal N} \rangle$ we find ${\rm max}_{\underline n \ge {\cal N}} |\langle \underline n | {\cal N}\rangle | $, which defines the place of $ | \underline n \rangle $ in the list, $ n = {\cal N}$. How this algorithm works on the truncated Hamiltonian, with omitted non-resonant matrix elements, is shown in Fig.~\ref{fig1}a) for the strong disorder $W=10$ and the $L=12$ system. The block-diagonal structure of $M_{\underline n,{ \cal N}}$ (and thus also the clustering) is fully apparent.

It remains to be checked that the clustering observed in Fig.~\ref{fig1}a) is not just an artifact of our truncation scheme in Eq.~(\ref{rc}). To this end, we order the exact eigenstates of the full Hamiltonian according to the maximal projections $|\langle \underline n | {\cal N}\rangle | $. For the basis states $\{| \underline n \rangle \}$ we keep the ordering obtained for the truncated Hamiltonian.  The result for the full Hamiltonian is shown in Fig.~\ref{fig1}b) for the same realization of disorder as in Fig.~\ref{fig1}a). Although some exact eigenstates exhibit non-vanishing overlaps with the basis states in multiple clusters, one sees that the block structure of $M_{\underline n,{ \cal N}} $ is present even when the RC is lifted. For the full Hamiltonian, we expect that an appropriately refined sorting may reveal an additional structure, e.g., multifractal properties \cite{serbyn15,serbyn17,roy191,mace19}.

\begin{figure}[t]
\includegraphics[width=1.0\columnwidth]{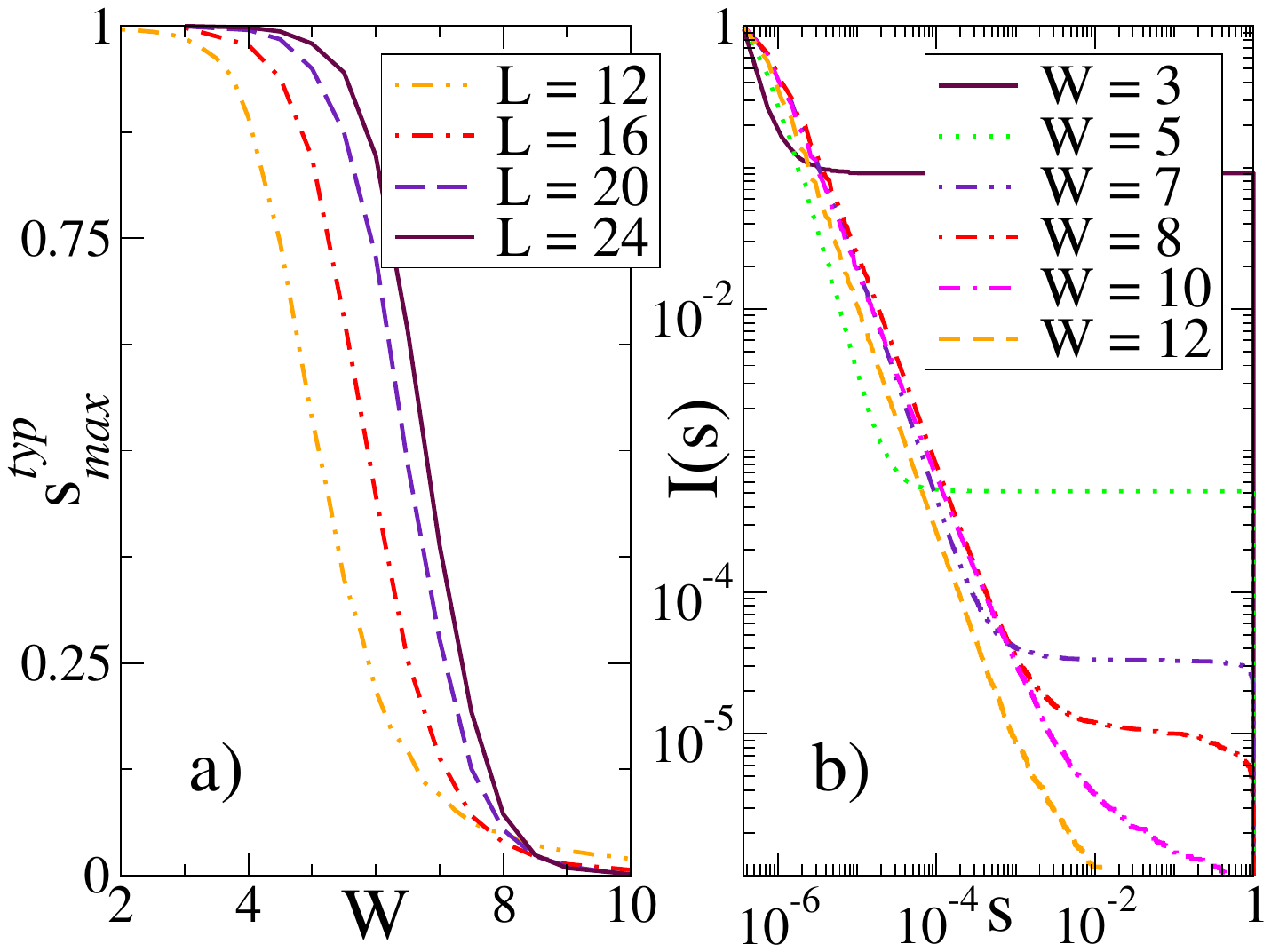}
\caption{a) $s^{typ}_{max}$ vs. disorder $W$, calculated for $R=0.3$ and $L = 12 - 24$, b) $I(s)$ for $W = 3 - 12$, evaluated for 
$L=24$ and averaged over $N_s=50$ samples.}
\label{fig2}
\end{figure}

Considering the RC explicitly, the connectivity problem and distributions of clusters in the Fock space may be directly studied, with the onset of macroscopic cluster(s) being the natural criterion for the breakdown of the MBL. We introduce $s=N_C/N_{MB}$, as the ratio of the cluster size, $N_C$, and the total number of MB states, $N_{MB}$, and look for universal behaviors related to the cluster sizes for the percolation type of transitions. We start by analyzing the $W$ dependence of the relative volume of the Fock space associated with the maximal MB cluster, $s_{max}$, evaluating typical averages, $s^{typ}_{max} = \exp {\langle \ln s_{max} \rangle_{dis} }$, with $\langle ...\rangle_{dis}$ representing the average over $N_{dis}$ different disorder configurations. The numerical results for $s^{typ}_{max}$ as a function of $W$ are shown in Fig.~\ref{fig2}a), involving sampling over $N_{dis} \geq 300$ disorder configurations. Although the results exhibit a pronounced $L$-dependence of $s^{typ}_{max}(W)$ for all $L$, it is important to notice from Fig.~\ref{fig2}a) that with increasing $L$ the drop in $s^{typ}_{max}(W)$ near the transition at $W_c \approx 8$ becomes sharper.

As a next step, we examine the probability distribution of MB clusters ${\cal P}(s)$, which in the $L\rightarrow\infty$ limit should be universal at the percolation transition $W \approx W_c$ \cite{kirkpatrick73,stauffer79,essam80}. In particular, in Fig.~\ref{fig2}b), we show the results in terms of the integrated distribution $I(s)= \int_s^1 {\cal P}(s') ds'$, using the normalization $I(s \to 0)=1$. The results in Fig.~\ref{fig2}b) are presented for the large system size $L=24$ ($N_{MB} \approx3\times10^6$), averaging over $N_{dis} \sim 50$ samples (realizations of disorder). For $W< W_c$, long plateaus exhibited by $I(s)$ in Fig.~\ref{fig2}b) are consistent with the existence of a single macroscopic cluster. That is, the long plateaus that reach $I(s=1)$ as a straight line in Fig.~\ref{fig2}b) can be obtained only if one large cluster, $s_{max}\approx 1$, exists for each realization of disorder. The bending of $I(s)$ near $s=1$ for larger $W$ indicates a degradation of the macroscopical cluster into a rapidly increasing number of small clusters as $W$ increases, with sample-to-sample fluctuations of $s_{max}$ due to finite $L$ near the percolation transition. Because of these fluctuations, the probability of finding a macroscopic cluster above $W_c$ remains finite, yet small (rare events), which is consistent with the very low values of $I(s\approx1)$ for $W\gtrsim 8$ in Fig.~\ref{fig2}b). 

Near the percolation transition, the distribution becomes universal in a broad range of $s$, exhibiting a power-law behavior ${\cal P}(s ) \propto s^{-\xi}$. A simple fit for $W = 8$ gives $\xi \sim 2.47$, which is very close to the value of $5/2$ predicted for percolation models in high dimensions ($D>6$) \cite{stauffer79,saberi15}.

A degradation of the macroscopic cluster may be related to the level statistics change under the increase of the disorder strength $W$, observed in previous studies \cite{oganesyan07,torres15,luitz15,serbyn16,suntajs19}. In this context, we analyze the standard quantity, i.e., the ratio of two consecutive level spacings, $s_{n+1}/s_n$, and the corresponding mean $r$ of  $r_n=\min(s_{n+1}/s_n,s_n/s_{n+1})$. For  strong disorder, the system is expected to exhibit the Poisson level distribution, with $r\approx0.386$. The larger value $r\approx0.536$ characterizes the Gaussian orthogonal ensemble (GOE) in the ergodic limit.

\begin{figure}[t]
\includegraphics[width=0.8\columnwidth]{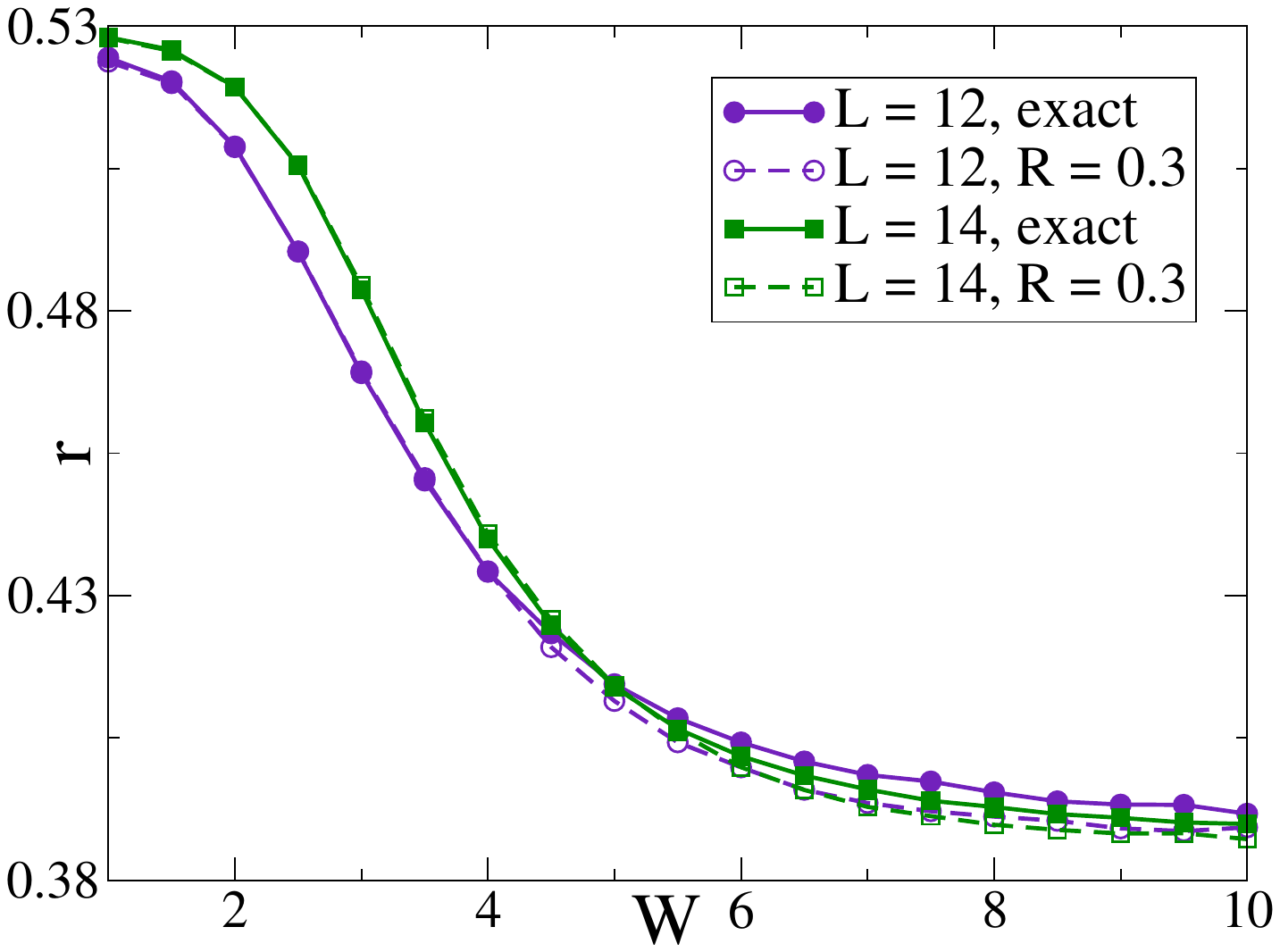}
\caption{Level statistics parameter $r$ as a function of $W$ reveals two limiting behaviors,  $r\approx0.536$ for small $W$ and $r\approx0.386$ for large $W$, corresponding to the GOE and the Poisson level distribution, respectively.}
\label{figLS}
\end{figure}

In Fig.~\ref{figLS}, the parameter $r$ is shown as a function of $W$, averaged over the spectra for $10^3$ different disorder samples, for small systems, $L=12, 14$, and open boundary conditions. For small and large $W$, all the curves clearly recover the limiting values of $r$ characteristic of the GOE and the Poisson level statistics, respectively. The exact results in Fig.~\ref{figLS} exhibit a smooth crossover between the two limiting behaviors. The only difference is a bit sharper crossover for $L=14$. The results obtained by omitting the non-resonant matrix elements ($R=0.3$) between clusters show almost the same behavior as the exact ones, with only small deviations for the strong disorder. The drop in $r$ is most pronounced for $2\lesssim W\lesssim5$, when most of the MB states are part of the same (degrading) macroscopic cluster.

\section{Charge-density relaxation}

\subsection{Rate-equation approach}

In standard bond percolation models the bonds may be switched on or off. In our case, each bond is different, given by the transition rate in Eq.~(\ref{re}). 
Our RE approach \cite{prelovsek18,mierzejewski19,mierzejewski20} in Eqs.~(\ref{re}) and (\ref{nnn}) offers an important advantage, allowing for the study of dynamical relaxations and the thermalization on the lattice starting from nonequilibrium MB states. In particular, we analyze relaxation properties of the charge-density profile,

\begin{equation}
\rho_q = \sum_{\underline n} n_q p_{\underline n}, \qquad
n_q = \sqrt{\frac{2}{L}} \sum_{il} \cos(q i)  |\phi_{li}|^2 n_l, \label{rhoq}
\end{equation}
for the smallest $q = \pi/L$ for systems with open boundary conditions. With the full diagonalization of $N_{MB}$ linear REs in Eq.~(\ref{re}), we can express the density correlation function (DCF) $C_q(t)= \langle \rho_q(t) \rho^0_{-q} \rangle$ in terms of eigenvalues $\lambda_\alpha$ and eigenfunctions $w_{\alpha \underline n}$, 
\begin{equation}
p_{\underline n} = \sum_\alpha c_\alpha w_{\alpha \underline n} \mathrm{e}^{-\lambda_\alpha t},
\quad \lambda_\alpha w_{\alpha \underline n} = - \sum_{\underline n} 
\tilde \Gamma_{\underline n \underline n'} w_{ \alpha \underline n'}\;.\label{eqlambdas}
\end{equation}
With a basis state as the initial nonequilibrium MB state, $p^0_{\underline n} = \delta_{\underline n \underline n_0}$, we get $c_{\alpha} = w_{\alpha \underline n_0}$. The averaging over all initial states yields,
\begin{equation}
C_q(t) = \frac{1}{N_{MB}} \sum_{\lambda_{\alpha} > 0} |W_{\alpha}|^2 \mathrm{e}^{-\lambda_{\alpha} t}, \quad
W_{\alpha} = \sum_{\underline n} n_q w_{ \alpha \underline n}\;, \label{cqt}
\end{equation}
where $\lambda_\alpha=0$ components are omitted, since they represent the 
equilibrium solutions, $p_{\underline n}= 1/N_{C}$, for each MB cluster. To discuss the anomalous diffusion, we 
introduce a time-dependent diffusion, given by,
\begin{equation}
D(t)= - (1/q^2) d [\ln C_q(t)]/dt\;. \label{dt}
\end{equation}

\subsection{Full quantum calculations}

As a reference for comparison with the approximate RE approach,  we first present  full quantum results for the dynamics within a quenched system.  In particular, we consider the Hamiltonian, Eq.~(\ref{tvh}), with periodic boundary conditions.  Initially, at time $t=0$,  instead of random local potentials, we introduce a harmonic perturbation  with the longest wave-length $\epsilon_i= \cos(2 \pi i/ L )$. Using the microcanonical Lanczos method, we find the initial state $| \psi(0) \rangle$ corresponding to the high but finite temperature, $1/k_BT\simeq 0.2$, and the small energy spread $\delta E \sim 10^{-3}$. Then, we quench the potentials $\epsilon_i$ with random values within $[-W,W]$, setting in addition  $\epsilon_{i}=\epsilon_{i-L/2}$. The initial state is propagated in time, $| \psi(0) \rangle \to  | \psi(t) \rangle$, and we evaluate the expectation values $\langle n_i(t) \rangle=   \langle \psi(t)| n_i | \psi(t) \rangle$.   The  spatial distribution of particles is then \mbox{$\langle n_i (t) \rangle=\sum_{q} \cos(q i) C_{q}(t)$}, whereby we focus on the smallest $q=2\pi/L$ and study the normalized Fourier component $\tilde C_{q}(t)=C_{q}(t)/C_{q}(0)$.  
For a diffusive system, one expects an exponential decay  \mbox{$\tilde{C}_{q}(t)=\exp(-D q^2 t)$}. However, we generally observe a time-dependent $D(t)$,
given by Eq.~(\ref{dt}). 

\begin{figure}[t]
\includegraphics[width=0.8 \columnwidth]{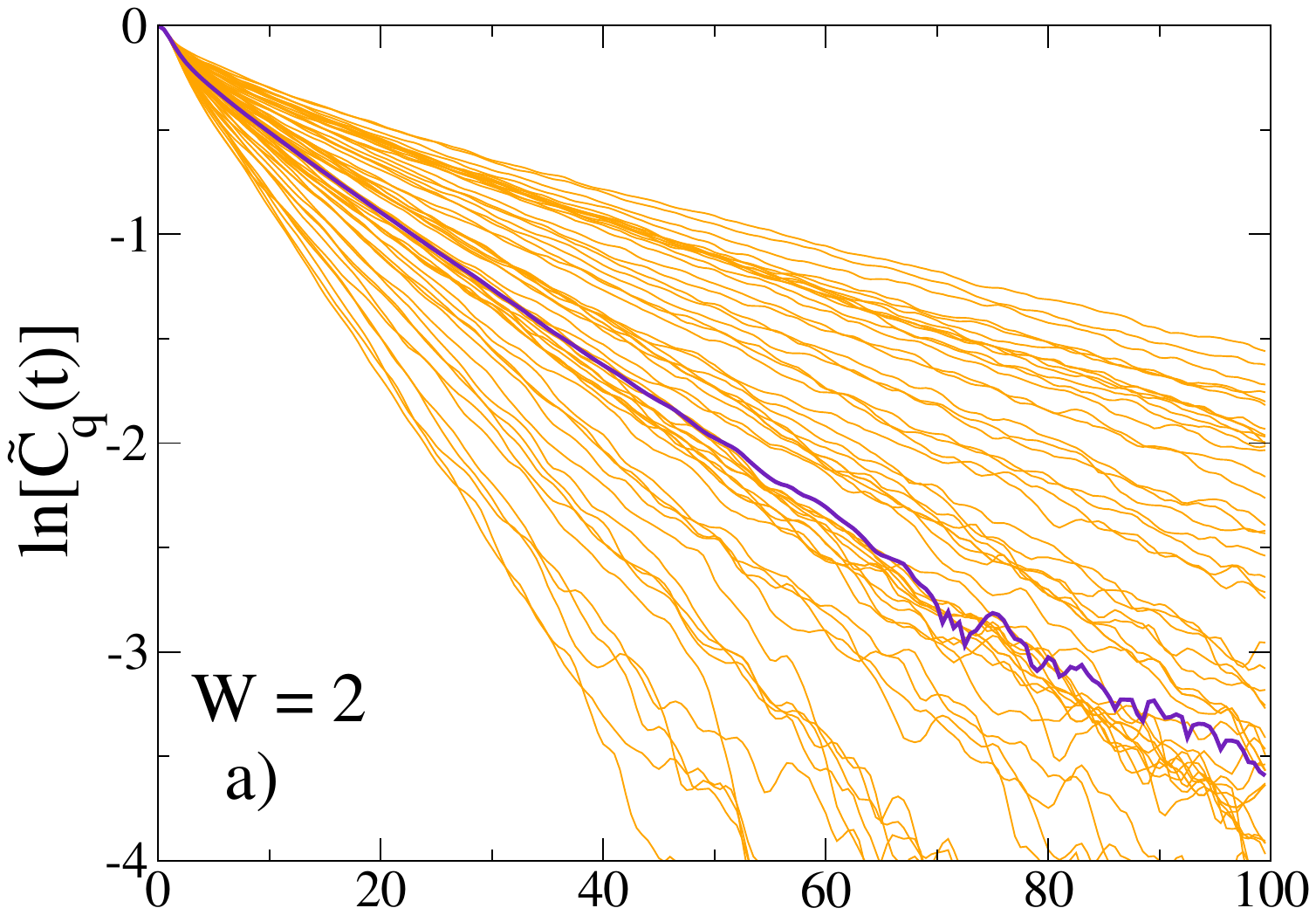}
\includegraphics[width=0.8\columnwidth]{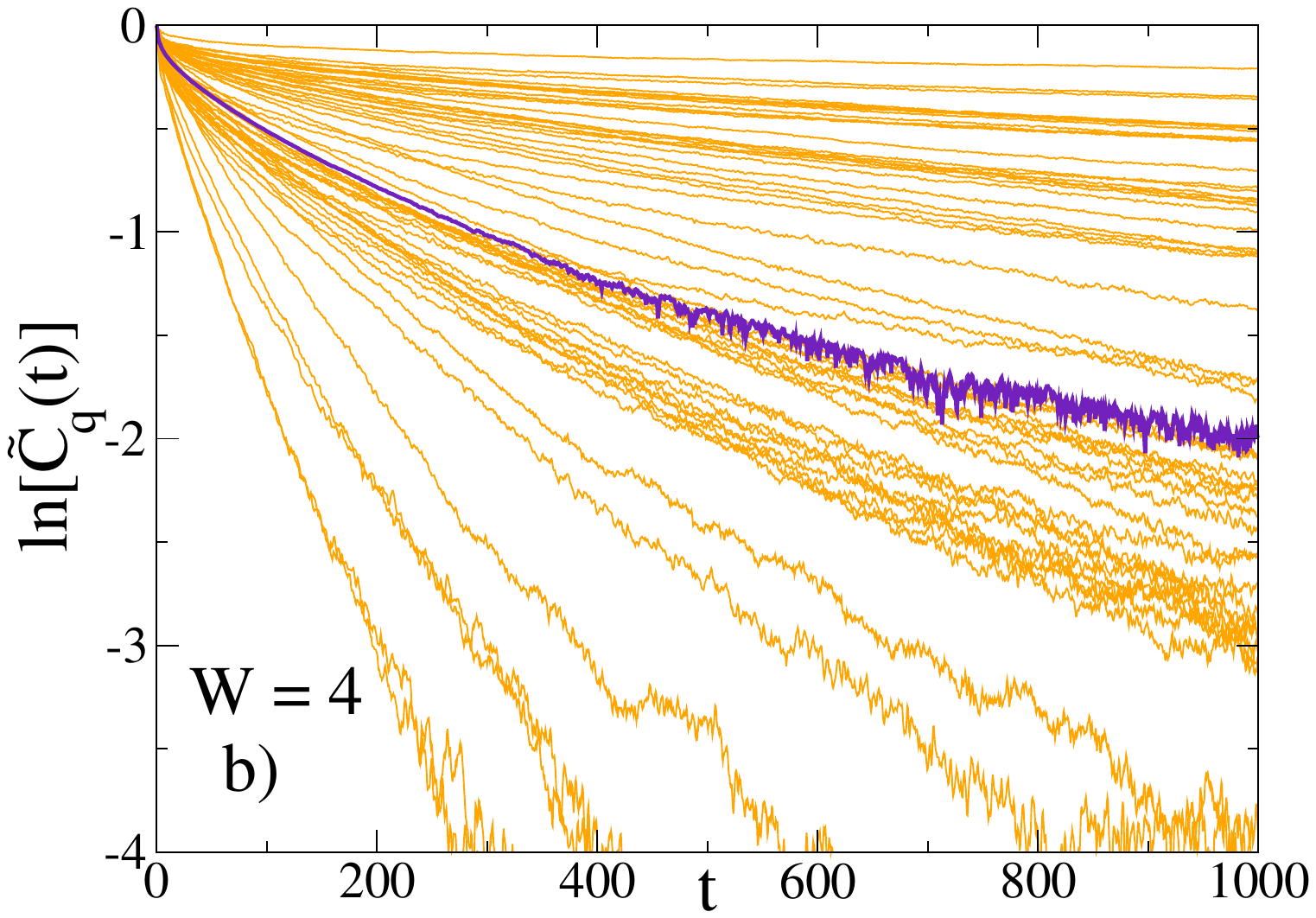}
\caption{Full quantum results for $\tilde{C}_q(t)$ for the chain with  $L=22$ sites and a) the disorder strength $W=2$ and b) $W=4$. Various thin curves correspond to different samples, while the thick line shows the typical value of $\tilde{C}_q(t)$.  }
\label{figS1}
\end{figure}

Various curves in Fig.~\ref{figS1} show $\tilde{C}_{q}(t)$ for different samples. In both cases in Fig.~\ref{figS1}, $W=2$ and $W=4$, the full quantum calculations clearly exhibit strong sample-to-sample fluctuations of the correlation function $\tilde{C}_{q}(t)$ . Due to the specific choice of random potentials ($\epsilon_{i}=\epsilon_{i-L/2}$), the Fourier transform of $\epsilon_{i}$ has no component with the smallest  $q$. This ensures that the numerical data are clear enough to obtain diffusion constants independently for each realization of disorder.  For $W=2$, we observe that all the samples are diffusive. Yet, the spread of $D$ is significant. For stronger disorders, $W=4$, this spread is even more pronounced. That is, for certain samples the system is nearly diffusive, while for some other samples, the inhomogeneous distribution of particles hardly changes in time. Thus, it may be concluded that with increasing $W$ the nature of the relaxation processes changes, and that even for large systems  the properties of these processes are very sensitive to a particular disorder realization. 

\subsection{Full diagonalization of the rate equations}

By exact diagonalization of the RE, one obtains the eigenvalues $\lambda_\alpha$ and eigenfunctions $w_{\alpha \underline n}$ in Eq.~(\ref{eqlambdas}). Each of these solutions corresponds to another RE relaxation process. Their character may be very different as a function of $W$, and in order to obtain better insight into this issue, it is particularly useful to calculate the inverse participation ratio for each of the solutions, IPR$_\alpha =\sum_{\underline n }| w_{\alpha \underline n}|^4$. IPR$_\alpha$ is a measure of localization of the solution in the basis of localized MB Anderson states. In the ergodic limit, one expects an IPR that is close to $1/N_{MB}$, whereas IPR $\approx1$ indicates an extreme opposite case when the system remains localized around one basis state.

\begin{figure}[tbh]
\includegraphics[width=0.8\columnwidth]{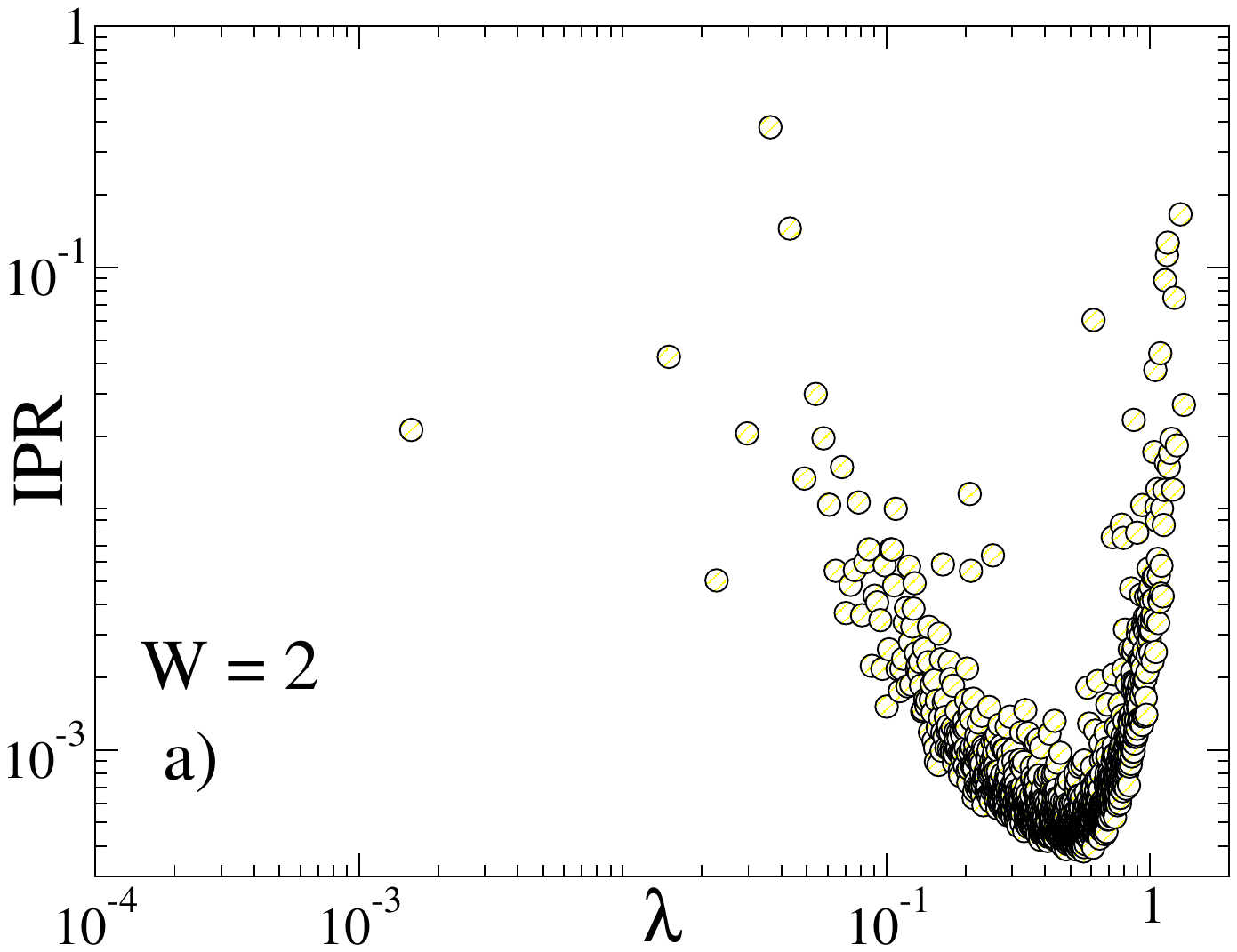}
\includegraphics[width=0.8\columnwidth]{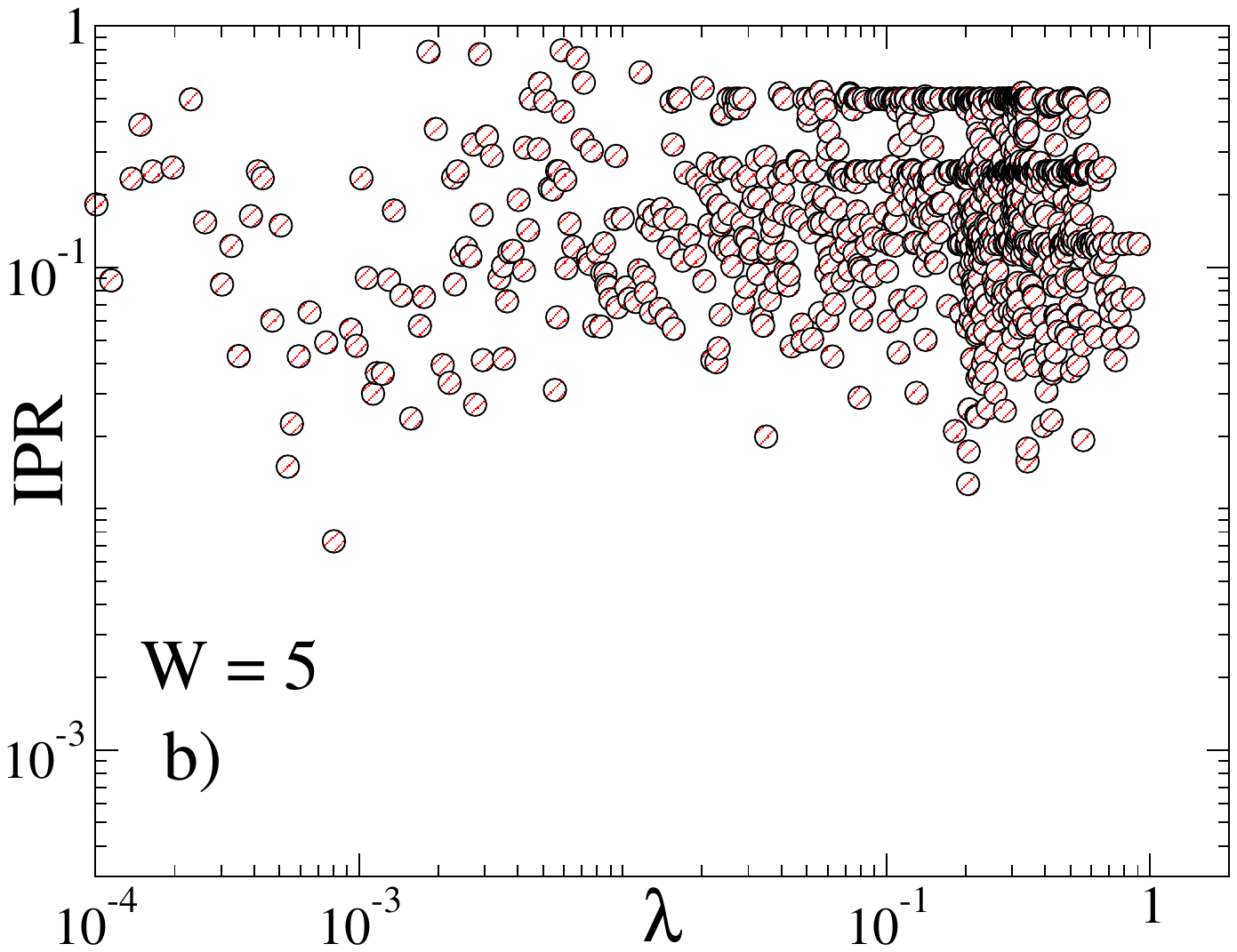}
\caption{The RE solutions for one configuration of disorder of the $L = 16$ system, shown in terms of points in the $\lambda$-IPR diagram. For the sake of clarity, every tenth point is plotted. a)  $W=2$, and b) $W=5$.}\label{figS2}
\end{figure}

For one disorder configuration and $L=16$, the exact solutions of the RE equations are shown in Fig.~\ref{figS2} in terms of [$\lambda_\alpha$, IPR$_\alpha$] points. A large $\lambda$ characterizes a fast relaxation, while a small IPR corresponds to a well delocalized mode in the MB space. In this respect, Figs.~\ref{figS2}a) and \ref{figS2}b), corresponding to the $W=2$ and $W=5$ cases, respectively, clearly describe fundamentally different regimes. In particular, with only a few exceptions, for $W=2$ all the relaxations processes are fast, with the great majority of these processes being characterized by small IPRs. On the other hand, for $W=5$, most of the eigenmodes are associated with the large IPRs.  Simultaneously, for $W=5$, the spreading of the points along the $\lambda$ axis is large, with some solutions exhibiting very small values of $\lambda$, representing very slow relaxation processes.  For $W=5$ in Fig.~\ref{figS2}b), it is interesting to note that some of the points are grouped along the horizontal lines, for which IPR $\approx1/2$, $1/4$, etc.  This behavior is apparently related to the formation of very small clusters in the Fock space.

\begin{figure}[t!]
\includegraphics[width=0.9\columnwidth]{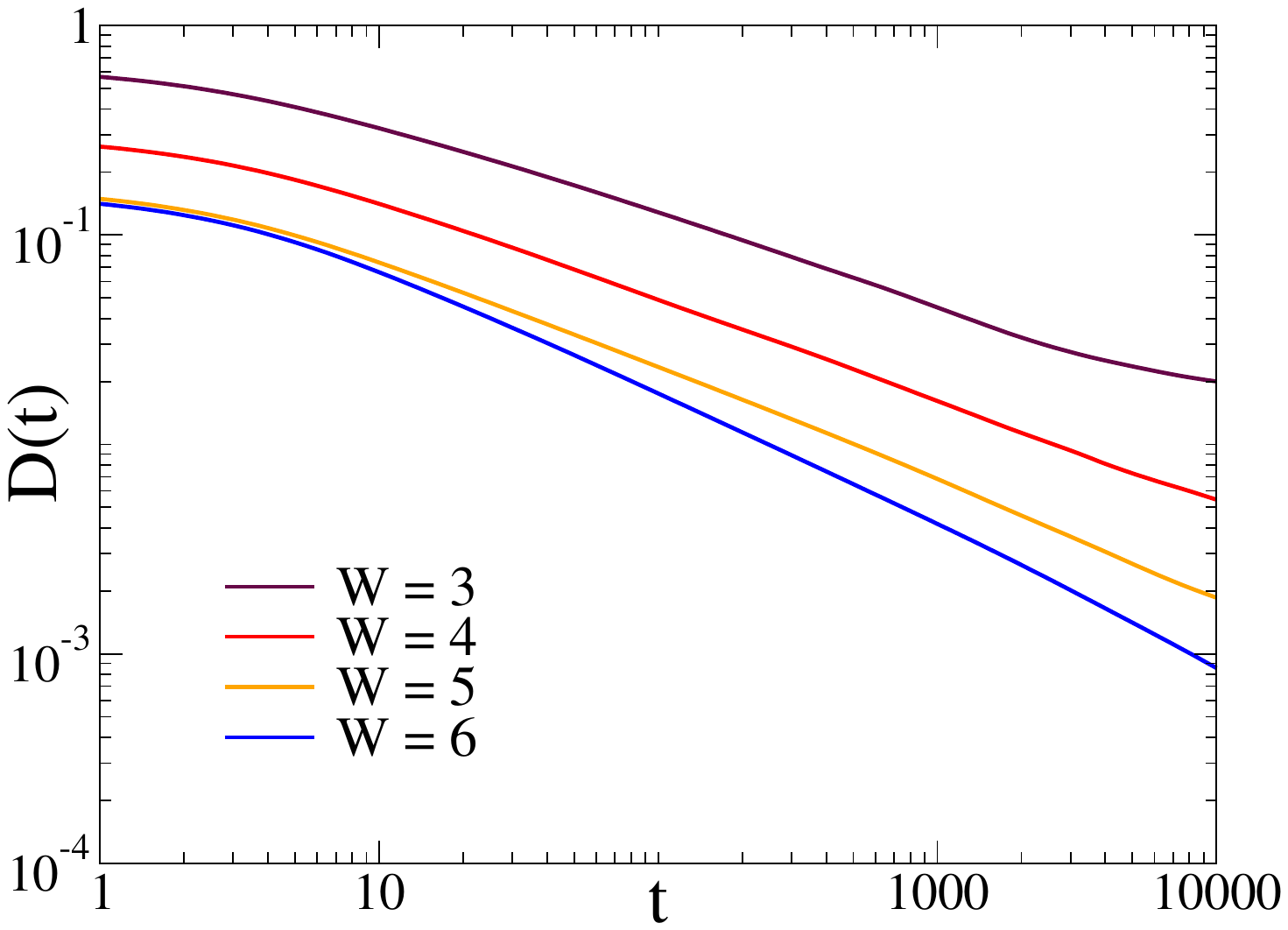}
\caption{Average $D(t)$ calculated by the exact diagonalization of the RE equations for $L=16$ and various $W = 3-6$. 
Averaging is performed over $N_{dis}=200$ configurations.}
\label{figS3}
\end{figure}

Due to the large number of MB states, $N_{MB}$, the full-diagonalization approach to the RE is restricted to smaller $L\leq 16$ systems. However, it allows us to evaluate the dynamics over long (or even extreme) times $t$. The properties of the time-dependent $D(t)$, given by Eq.~(\ref{dt}), may be seen from Fig.~\ref{figS3}. The results are obtained in the intermediate regime $3< W<W_c$, beyond the regime of normal diffusion $W> W^* \sim 2$. The curves represent the average $D(t)$, calculated for $N_{dis} = 200$ samples, by including in the average all the MB clusters.

For the considered $W\gtrsim 3$, even for the very long times in Fig.~\ref{figS3}, many eigenmodes contribute to $D(t)$. As the faster relaxation  processes die out, corresponding to larger $\lambda_\alpha$, $D(t)$ monotonically decreases.  Thus, with the increasing disorder, the system remains for the largest times 
in Fig.~\ref{figS3} with only slower relaxation processes still active. This latter behavior is related to the distribution of [$\lambda_\alpha$, IPR$_\alpha$] points in Fig.~\ref{figS2}, exhibiting a large spreading of values of $\lambda_\alpha$ over multiple orders of magnitudes.

\subsection{Direct time integration of the rate equations}

For shorter $t \leq 10^3$, we employ the direct time integration of the RE, Eq.~(\ref{re}), allowing for the study of large systems up to $L = 24$. To avoid infeasible sampling of the DCF over all initial states $|\underline n\rangle$, a convenient initial  inhomogeneous distribution is chosen, $p_{\underline n}(t=0) \propto \exp[- \hat n_q p_{\underline n}/T]$, corresponding to a potential imposing an initial charge-density modulation ($T = 0.5$). A modified DCF is calculated next, $\tilde C_q(t) = \langle \rho_q(t) \rangle/\langle \rho^0_q \rangle$. Since the macroscopic MB cluster should provide the dominating contribution to the ergodicity and transport properties across the whole system, in the following we analyze only the latter.

\begin{figure}[t!]
\includegraphics[width=0.9\columnwidth,angle=0]{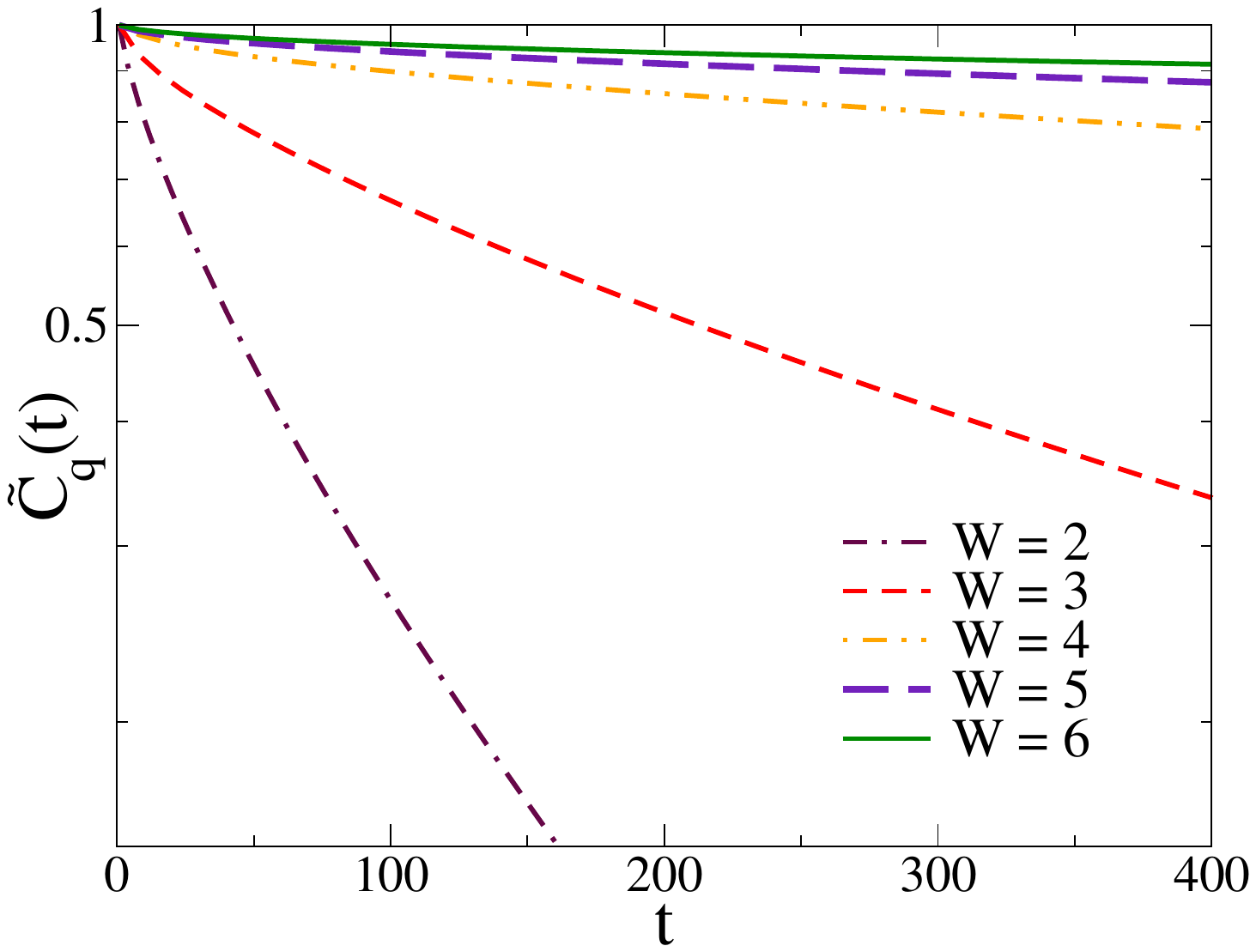}
\caption{Typical DCF $\tilde C_q(t)$ for the $L=24$  system and
various $W = 2 - 6$, obtained via time integration of the RE.}
\label{fig3}
\end{figure}

\begin{figure}[tb]
	\includegraphics[width=0.8\columnwidth]{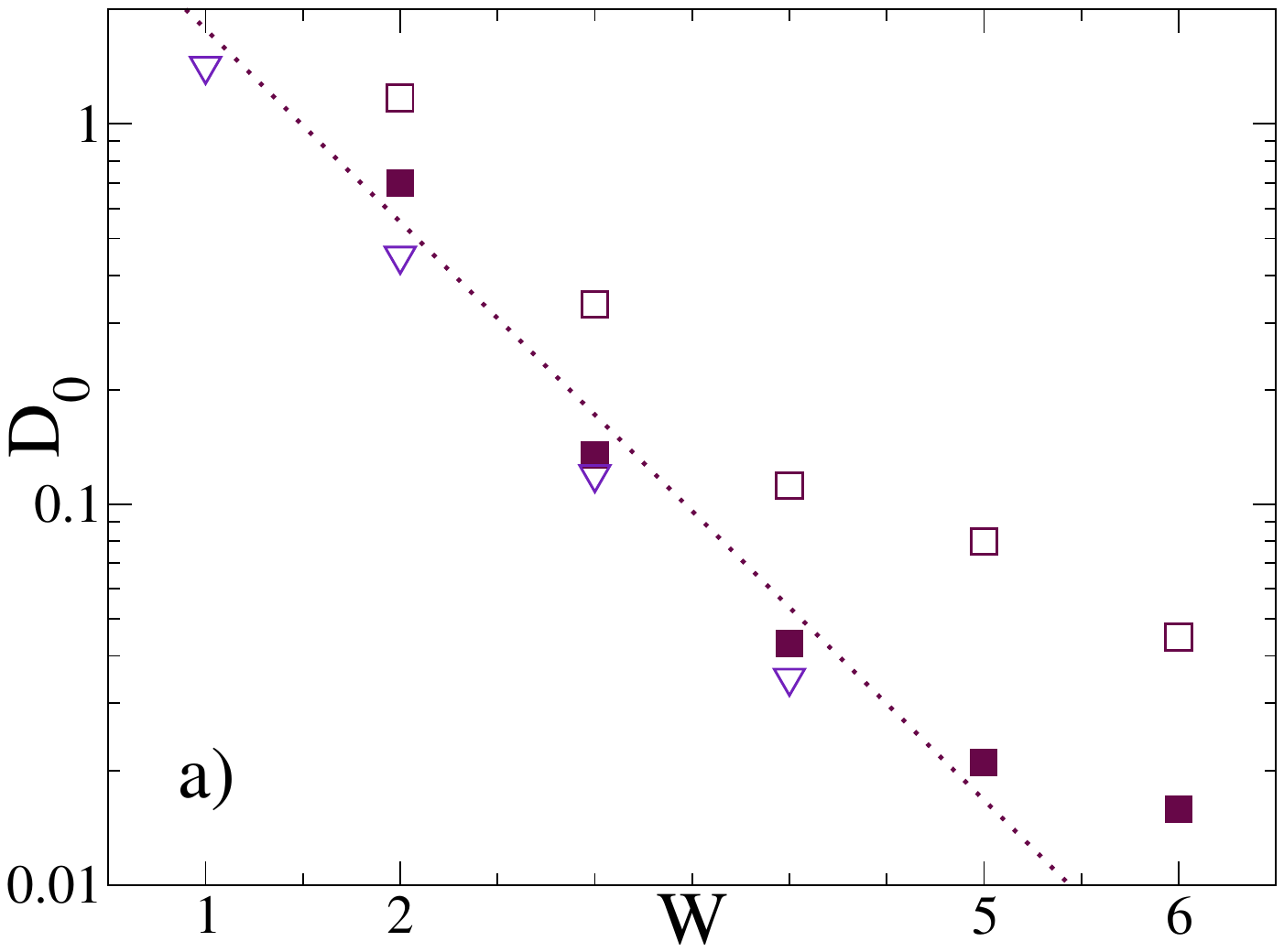}
	\includegraphics[width=0.8\columnwidth]{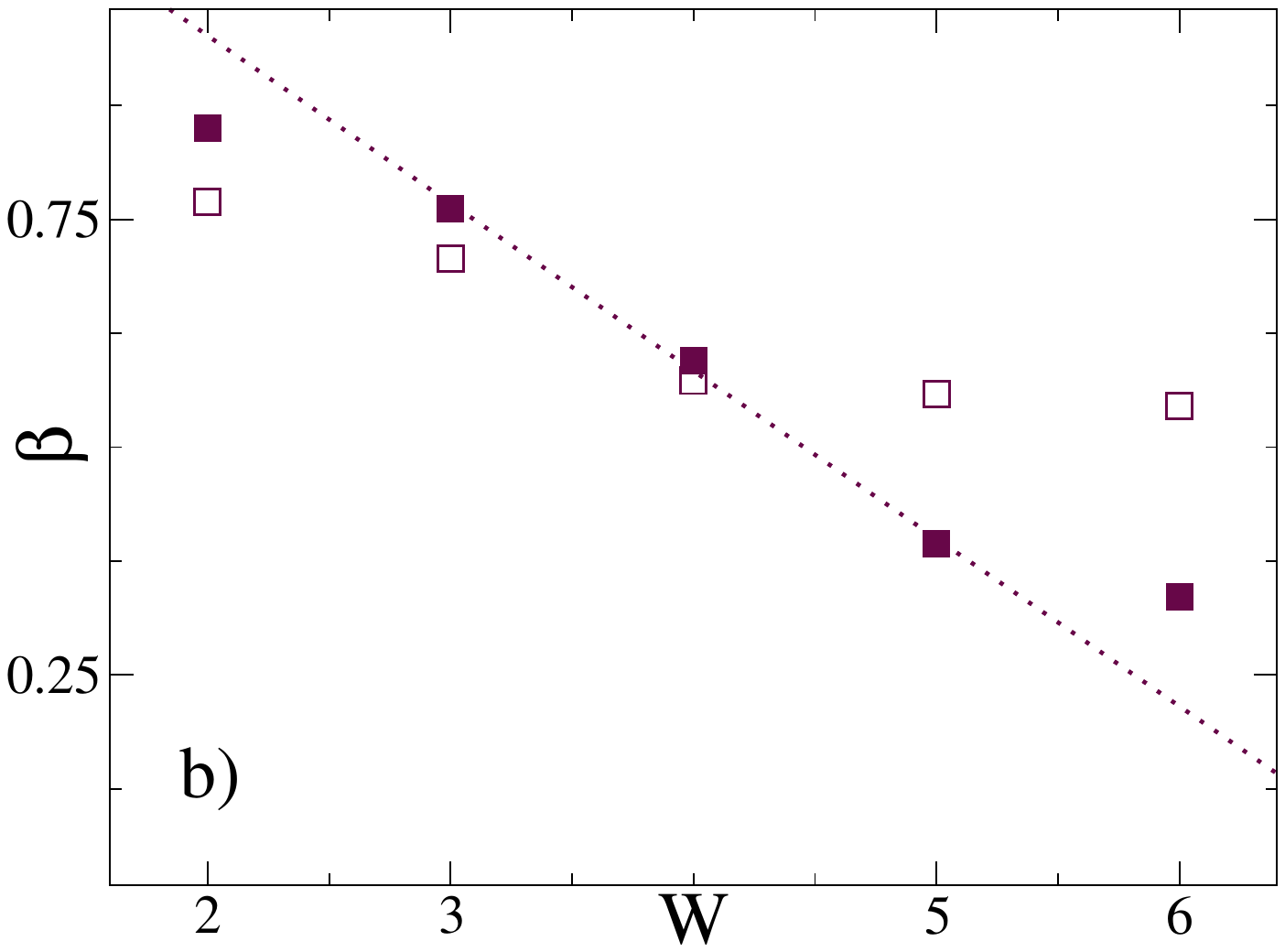}
	\caption{a) Effective diffusion parameter $D_0$ and b) the exponent $\beta$. Open squares show results from
		typical $\tilde C_q(t)$ for $L=24$ on Fig. \ref{fig3}, and solid squares show results from the  median $D_m(t)$ as in Figs. \ref{fig4}a) and \ref{fig4}b). 
		In a) we also plot $\tilde D$ (open triangles) from the full quantum calculation for $L=22$.}
	\label{fig5}
\end{figure}

In Fig.~\ref{fig3}, typical $\tilde C_q(t)$ are shown for $L=24$ and $N_{dis} = 50$. For $W>2$, it is rather evident that the DCF is not just a simple exponential one, expected for normal diffusion, $D(t)=$ const. Indeed, the decays in Fig.~\ref{fig3} are much better represented by a stretched-exponential form, $\tilde C_q(t) = \exp[  - D_0 q^2 t \; A(t) ]$. Here $D_0$ is an effective diffusion parameter, while $A(t)=\frac{1}{\beta} (t_0/t)^{1-\beta}$ introduces  corrections due to non-diffusive dynamics for $\beta <1$. In order to get $A(t)$  of the order of unity for the studied time window, $t\sim 10^2$, we set $t_0=50$.
Values of $\beta$ are plotted in Fig.~\ref{fig5}b) (open symbols), ranging from $\beta = 0.76$  for $W=2$ to $\beta = 0.54$ for $W=5$, 
whereas  $D_0$  is exponentially suppressed with $W$, as shown by open symbols in Fig.~\ref{fig5}a). 
 These charge-modulation decays may be rationalized by associating the behavior of $\beta$ to the sparsening of the MB connectivity graph while relating the behavior of $D_0$ to the overall decrease of transition rates in Eq.~(\ref{re}).

\begin{figure}[tb]
	\includegraphics[width=1.0\columnwidth]{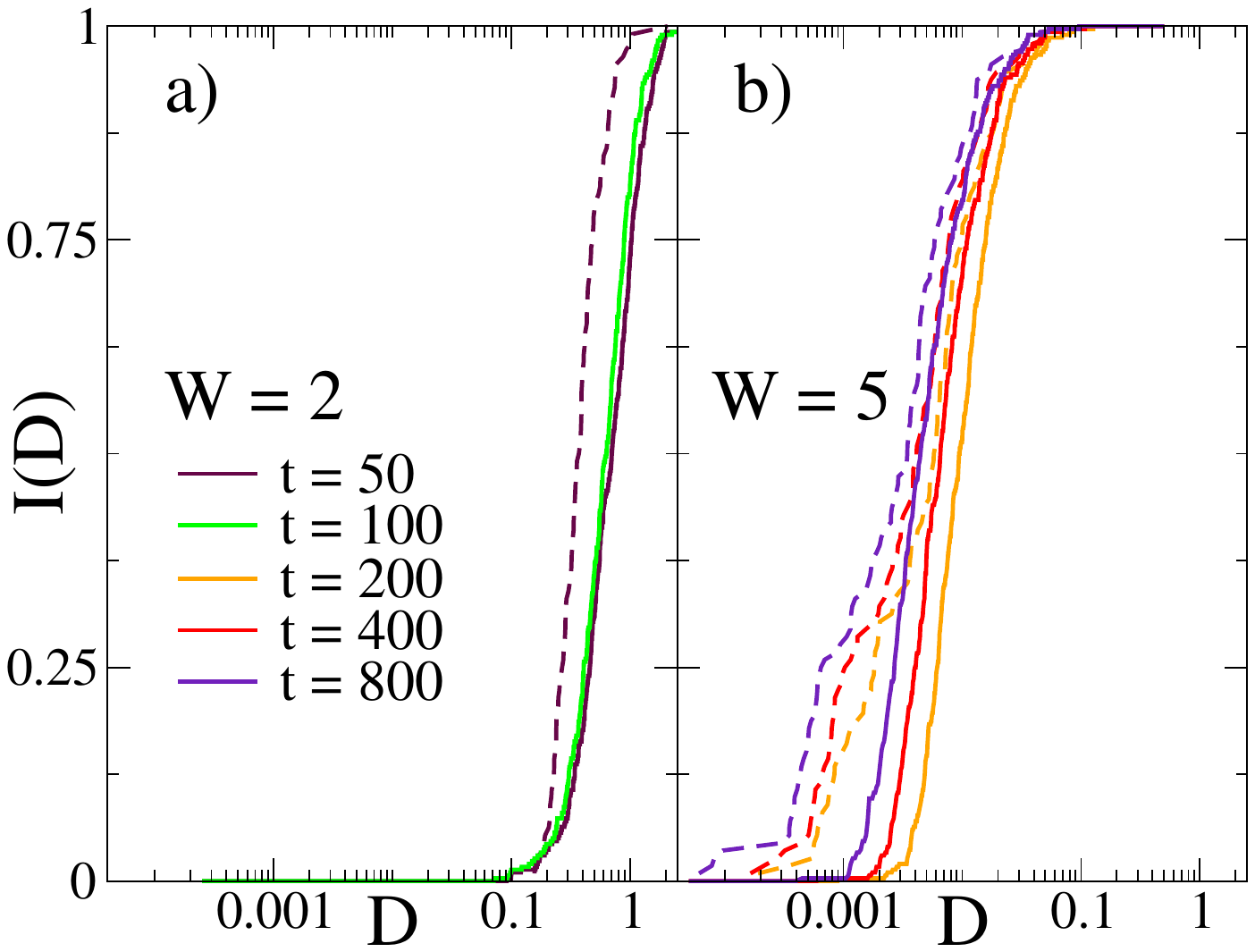}
	\caption{Integrated distribution of diffusion parameter $I(D)$, evaluated via the RE for $L=24$ at different
		times $t$, for two different disorders: a) $W =2$ and b) $W=5 $. Also presented are corresponding 
		full quantum results (dashed curves), obtained for $L=22$.}
	\label{fig4}
\end{figure}

Since $\tilde C_q(t)$ in Fig.~\ref{fig3} is a (typical) averaged quantity, more complete insights may be provided by calculating $D(t)$ in Eq.~(\ref{dt}) for each disorder configuration separately and by studying how $I(D)$, the (integrated) distribution of $D(t)$, evolves in time. For different times, $t = 2^n t_0$, $I(D)$ is shown  in Figs.~\ref{fig4}a) and \ref{fig4}b) for $W=2,5$, respectively,  obtained for the $L=24$ system and $N_{dis} = 300$. Furthermore, for comparison, we present in Fig.~\ref{fig4} results of the fully quantum time evolution of $I(D)$ as well. The latter is calculated for $L=22$. For both cases, $I(D)$ reveals a broad distribution of $D$. In particular, the RE results appear to follow roughly  a log-normal distribution, which is symmetric on the $\ln D$ scale around the median $D_m$, defined by $I(D_m)=1/2$. For $W=5$ and small $D < 10^{-3}$ in Fig.~\ref{fig4}b), discrepancies between the two approaches may be observed, which might originate from the fact that the quantum result includes the whole system, while the RE includes only the macroscopic clusters. 

A quite uniform shifting of $I(D)$ along the logarithmic time scale in Fig.~\ref{fig4}b) is fully consistent with the stretched-exponential decay of DCF, namely, $D_m(t) \sim D_0 (t_0/t)^{1-\beta}$, again with $\beta<1$. 
It should be noted that for the disorder $W=2$, as $L$ increases, we get $\beta \to 1$, and the normal diffusion $D(t)= D_0=$const is recovered. We expect that the most reliable values of $\beta$ and $D_0$ are obtained by considering the median $D_m(t)$. Namely, because of the very broad nature of $I(D)$ distributions, it is apparent that the estimate based on the typical $\tilde C_q(t)$ is obscured by the strong sample-to-sample fluctuations.

Figure~\ref{fig5} summarizes the results of the RE approach for parameters $D_0$ and $\beta$, characterizing the anomalous transport behaviors in the transient regime, $2 < W < W_c$.  The overall dependency of $D_0$ observed from Fig.~\ref{fig5}a) is exponential, i.e., $D_0 \propto \exp(- b W)$, with $b \sim 1.2$, while $\beta$ is approximately linear in $W$. In Fig.~\ref{fig5}a), the effective $\tilde  D$, obtained with the full quantum calculations, is plotted by open triangles, exhibiting exponential dependence on $W$ as well. Here we should  recall the relation between $\tilde D$ and dc conductivity $\tilde \sigma_0$, the latter being calculated for the same model in previous studies  \cite{barisic10,prelovsek17}, where at $T \gg W$ one gets $\tilde \sigma_0 = T\sigma_0 = \tilde D/4$, and, essentially, the very same coefficient, $b \approx 1.1$ \cite{prelovsek17}. This gives additional justification of the RE approximation and the overall physical picture that follows from it in terms of percolative clustering in the MB space and very different relaxation processes within clusters.

\section{Conclusions}
 
We have demonstrated that eigenstates of the strongly disordered MB system form cluster like structures, provided that
in the present model eigenfunctions are expressed in the MB Anderson states. This effect 
becomes even more evident by considering only MB matrix elements satisfying the resonant criterion. The emerging connectivity problem 
reveals the percolation threshold at the critical disorder $W_c$, characterized by the disappearance of the macroscopic cluster 
and the universal distribution of cluster sizes. Within our approximate RE approach, the location of $W_c$  depends on the resonance parameter $R$,
which gives for the chosen $R=0.3$ a very reasonable estimate, $W_c \sim 8$, consistent with the full quantum
calculations. 

The charge relaxation (diffusion)  studied in the context of the RE below the MBL transition even quantitatively agrees with the full quantum results, 
with the exponential suppression of the effective diffusion constant with $W$. The diffusion is anomalous
in the wide disorder range, $2 < W < W_c$, being characterized by the dynamical exponent $ \beta  < 1$. The obtained anomalous diffusion indicates a 
weakly coupled MB substructure within the degrading largest/macroscopic cluster, but with no evident/simple relation to the weak-link scenario in real space.

It should be acknowledged that in this work we did not consider the possible influence of higher-order resonances. They might affect the location of the MBL transition, plausibly shifting $W_c$ to higher values (analogous to reduction of $R$). Thus, we cannot exclude even the scenario of unbounded $W_c$, which would turn the transition into the glass-type crossover. Still, it is not expected that this would qualitatively change the overall description of the relaxation within the intermediate anomalous-diffusion regime.

\begin{acknowledgments}
P.P. acknowledges the support of the Project No. N1-0088 of the Slovenian Research Agency.
J.K. and O.S.B. acknowledge the support from Croatian Science Foundation Project No. IP-2016-06-7258. 
O.S.B. acknowledges the support from the QuantiXLie Center of Excellence, a project co-financed by the Croatian
Government and European Union through the European
Regional Development Fund - the Competitiveness and Cohesion
Operational Program (Grant No. KK.01.1.1.01.0004).
M.M.  acknowledges support from the National Science Centre, Poland via Project No. 2016/23/B/ST3/00647.
\end{acknowledgments}

\bibliography{manumblpc}
\end{document}